\documentclass[a4paper,11pt]{article}
\usepackage{aaskaiid}
\usepackage{orcidlink}
\usepackage{blindtext}
\usepackage{tcolorbox}
\usepackage{graphicx}

\title{Gamma-ray Bursts and Kilonovae from Gravitational Wave Events }
\ShortTitle{GRBs and KNe from GW events}

\author[1,2]{Alberto Colombo\orcidlink{0000-0002-7439-4773}$^\dagger$}
\ShortName{Colombo et al.} 
\author[3]{Marcello Giroletti\orcidlink{0000-0002-8657-8852}$^\dagger$}
\author[4]{Susanna Vergani\orcidlink{0000-0001-9398-4907}$^\dagger$}
\author[5,6]{Lauren Rhodes\orcidlink{0000-0003-2705-4941}}
\author[]{the SKA Transients SWG}

\affiliation[1]{INFN -- Sezione di Roma, 1-00185 Roma, Italy}
\affiliation[2]{INAF -- Osservatorio Astronomico di Brera, via Emilio Bianchi 46, I-23807 Merate (LC), Italy}
\emailAdd{alberto.colombo@inaf.it}
\affiliation[3]{INAF -- Istituto di Radioastronomia, via P. Gobetti 101, 40129, Bologna, Italy}
\emailAdd{marcello.giroletti@inaf.it}
\affiliation[4]{LUX, Observatoire de Paris, Université PSL, CNRS, Sorbonne
Université, Meudon, 92190, France}
\emailAdd{susanna.vergani@obspm.fr}
\affiliation[5]{Trottier Space Institute, McGill University, 3550 rue University, Montr\'eal, QC H3A 2A7, Canada}
    \affiliation[6]{Department of Physics, McGill University, 3600 rue University, Montr\'eal, QC H3A 2T8, Canada
\\\vspace{0.5cm}  $^\dagger$ These authors contributed equally to this work and are corresponding authors.}

\abstract{
The detection of gravitational waves (GWs) from binary black holes in 2015 and the joint GW–electromagnetic (EM) observation of the binary neutron star merger GW170817 set a milestone in the multimessenger era in astrophysics. After four observing runs by the LIGO, Virgo, and KAGRA interferometers, a new cycle is planned for 2028, paving the way for next-generation detectors in the 2030s—such as the Einstein Telescope, Cosmic Explorer, and LISA.
The prospects for joint GW–EM studies, including kilonova searches in wide optical surveys, are vast but demanding. In the radio domain, connected interferometers and VLBI arrays have already proven essential in constraining the ejecta properties of GW170817. Radio emission from gamma-ray burst (GRB) afterglows, whether on- or off-axis, remains detectable for very long time, making radio observations the most effective method for identifying and tracking GW merger counterparts. These observations enable precise characterization of system evolution, detailed probing of GRB jet structures, and possible detection of misaligned jets once their velocity becomes non-relativistic.
Even in its initial configuration (AA*), the SKAO will provide the sensitivity and field of view needed to complement GW counterpart searches during O5 and beyond, offering unmatched capabilities for long-term monitoring. 
Furthermore, independent of the GW detections, SKAO will enable population studies of the properties of both long (produced by the collapse of massive stars) and short (produced by the merger of neutron stars) GRBs, of their jets and of their environment.
We present an overview of this evolving observational landscape and of the key scientific questions SKAO will address.}

\begin{document}
\maketitle

\section{Introduction}

Gamma-ray bursts (GRBs) are powerful ultra-relativistic jets associated with the collapse of massive stars or with the merger of compact objects. Their characteristics make them exceptional laboratories to study physical processes in extreme conditions and they are also used as tools to explore galaxies and the high-redshift universe.

The detection of the binary neutron star (BNS) merger GW170817 and its GRB electro-magnetic counterpart \citep{2017PhRvL.119p1101A,2017ApJ...848L..12A}, with its associated afterglow and kilonova emissions, put GRBs at the forefront of multi-messenger astrophysics. Furthermore, we are living a very exciting era where different GRB missions are operating at the same time. In particular, the new SVOM \citep{2016arXiv161006892W} and Einstein Probe \citep{2025SCPMA..6839501Y} missions are opening a discovery space as their instruments extend to a larger energy range the exploration of GRBs and afterglows from space. 

\begin{figure}[h]
    \centering
	
\includegraphics[width=0.7\columnwidth]{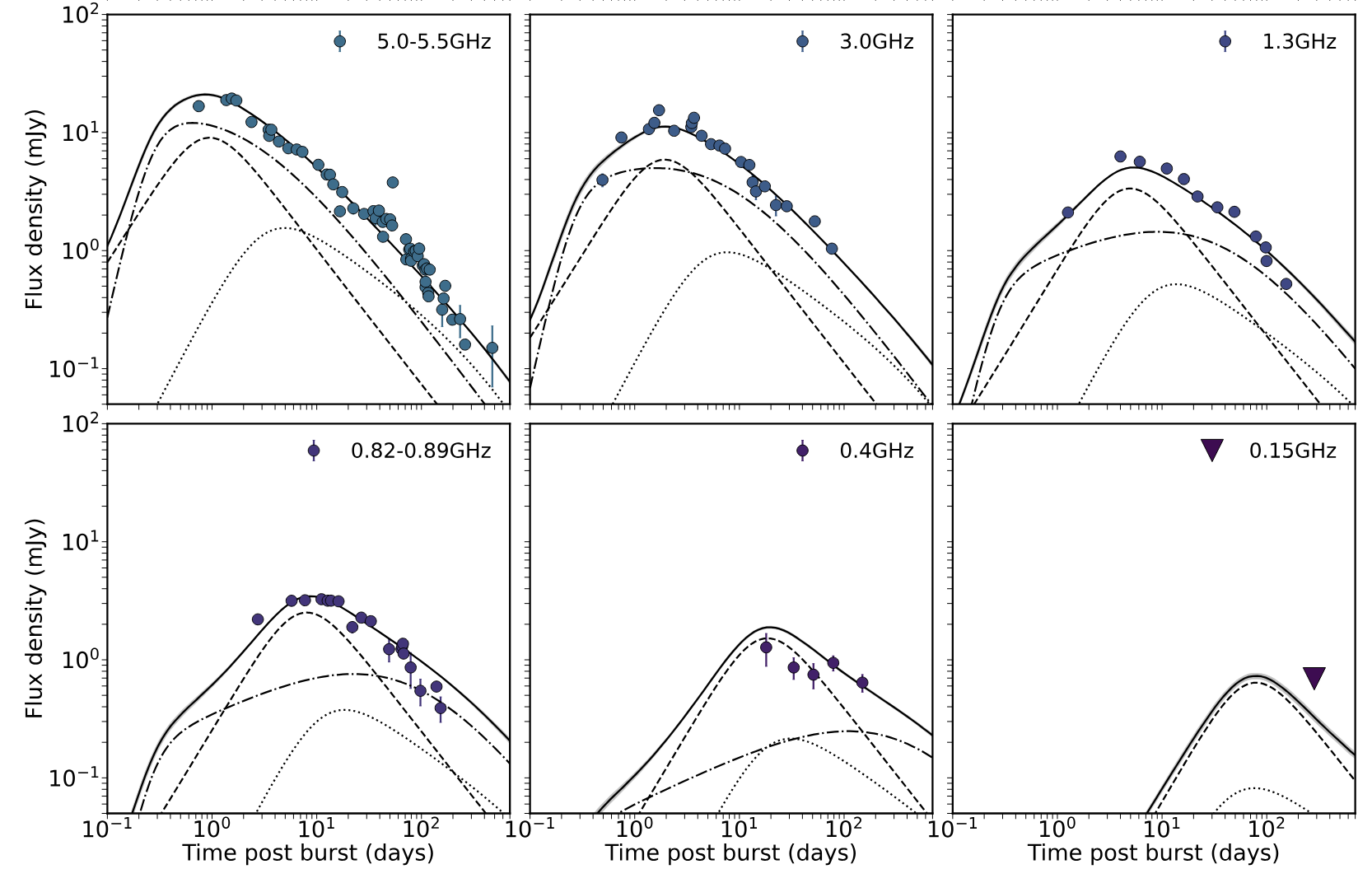}
\includegraphics[width=0.7\columnwidth]
    {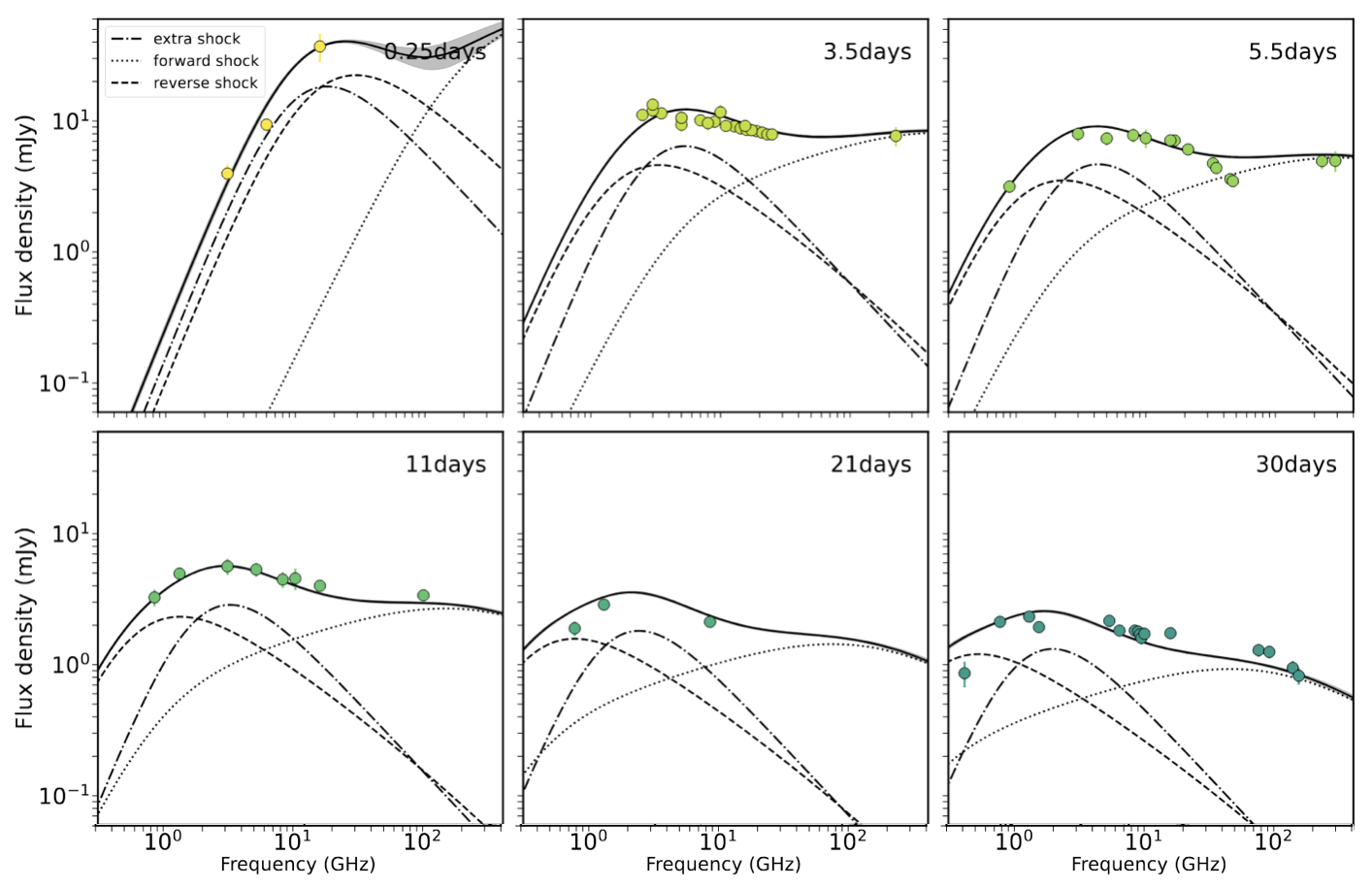}
    \caption{Part of the light curves (top) and spectral energy distribution (bottom) of long-duration GRB 221009A \citep{2024MNRAS.533.4435R}. Radio long timescale and multi-frequency observations were vital in order to deconstruct the afterglow into three separate shock components originating from a forward shock, reverse shock and broader cocoon as denoted by the dot-dash, dashed and dotted lines, respectively.}
    \label{fig:rhodes24}
\end{figure}

Studies of GRBs in the radio domain provide complementary and sometimes unique diagnostics on GRB explosions and their environments. Radio observations are fundamental to determine the emission processes at play (e.g.: reverse shock vs forward shock), the energetics of the jet, the jet structure and expansion, as well as the structure and density of the medium surrounding the GRB (e.g.: \citealt{2008A&A...480...35V}, \citealt{2017JApA...38...56R} and references therein, \citealt{2021MNRAS.503.1847L}; Figure \ref{fig:rhodes24}). The radio emission has the advantage (compared to optical) of peaking on timescales of a few days and generally staying bright up to several tens of days after the burst. Furthermore, observations at radio frequencies can also contribute significantly to the detection of dust-obscured or high-redshift GRB afterglows, whose optical emission is totally or partly absorbed (e.g.: \citealt{2022ApJ...940...53S}). 

\begin{figure}[h]
    \centering
	\includegraphics[width=0.4\columnwidth]{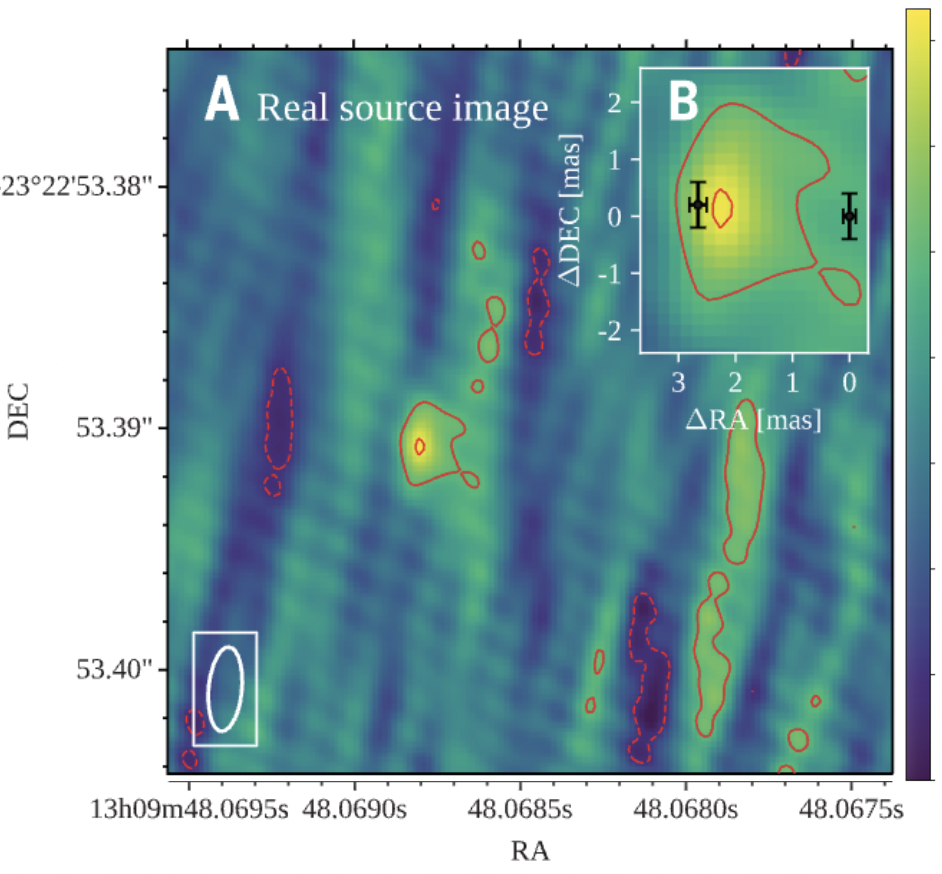}
\includegraphics[width=0.4\columnwidth]{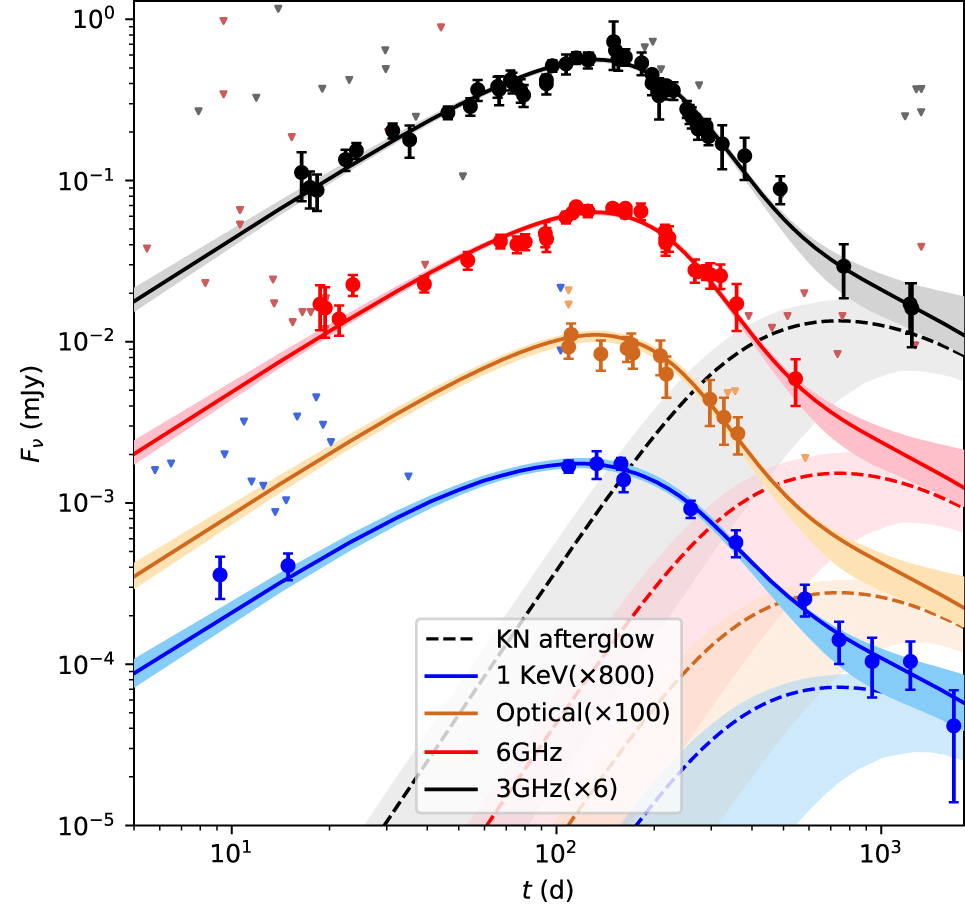}
    \caption{\textit{Left Panel (adapted from \citealt{2019Sci...363..968G}}): (A): Global-VLBI observed radio images of GRB\,170817A. The apparent source size is constrained to be smaller than 2.5\,mas at the 90\% confidence level. (B): a zoom on the position of the source, with black error bars showing  positions at 75 days and 230 days after the merger \citep{2018Natur.561..355M}, probing the superluminal motion of the source. \textit {Right Panel (adapted from \citealt{2023ApJ...943...13W}}): radio, optical and X-ray afterglow data from 5 to 1800 days after GW\,170817/GRB\,170817A. The dashed lines show the possible but uncertain kilonova afterglow contribution to the data interpretation.
    The light-colored regions represent the 90\% credible regions of the estimated kilonova light curves.}
    \label{fig:knaft}
\end{figure}

The GRB emission is beamed in the direction of the jet motion. To observe the prompt GRB emission, our line-of-sight has to lie within the jet cone. However, as the jet gradually decelerates over time, the beaming angle of the emission becomes wider, allowing the afterglow detection even for an observer not placed along the jet cone. In fact, on-axis GRBs are only the tip of the iceberg of the GRB population. For each GRB seen on-axis, there should be hundreds of GRBs for which only the off-axis afterglows can be detected \citep{2015A&A...578A..71G}. 
The first robust detection and detailed study of an off-axis afterglow was possible thanks to the multi-messenger campaign triggered by the detection of GW170817. Radio observations were fundamental in assessing the nature of the EM counterpart detected at late time as the off-axis afterglow of GRB170817 \citep{2017Sci...358.1579H,2018Natur.554..207M,2018ApJ...858L..15D}. Furthermore, global Very Long Baseline Interferometry (VLBI) observations allowed for the first time the study of the structure of the jet and the determination of its size (\citealt{2019Sci...363..968G}, Figure \ref{fig:knaft}) and apparent superluminal motion \citep{2018Natur.561..355M}. Radio data allowed for the refinement of the measurement of the inclination angle of the GW detection \citep{2022Natur.610..273M}, allowing for a much better determination of H$_0$ from GW events \citep{2019NatAs...3..940H}. Radio observational campaigns are still ongoing to look for the kilonova afterglow emission, but for the moment they are inconclusive (\citealt{2021ApJ...914L..20B}; Figure \ref{fig:knaft}). 

In the previous Science Book for the SKA, the motivation for observations of GRBs was described in detail, including measuring energy budgets and surrounding density profiles,  detecting orphan afterglows from off-axis jets, and tracking afterglow emission through the non-relativistic transition \citep{2015aska.confE..52B}. Since that publication, the detection of GWs and especially of GW~170817 has represented the main element of novelty. Therefore, we focus this Chapter primarily on the potential of the SKAO for the study of GW counterparts in a multi-messenger context (Sect.~\ref{s.2}) followed by an update on the scientific impact of the SKAO for GRBs in general (Sect.~\ref{s.3}); the Chapter is then concluded by an executive summary.

\section{The multi-messenger observational landscape}\label{s.2}

The current network of second-generation (2G) GW detectors — Advanced LIGO \citep[aLIGO,][]{aasi2015}, Advanced Virgo \citep[aVirgo,][]{acernese2015}, and KAGRA \citep{aso2013,abbott2018} — has opened the era of GW astronomy, with over two hundred binary black hole (BBH), two BNS
and several black hole - neutron star (BHNS) mergers detected to date \citep{2025arXiv250818082T}. In the near term, the addition of LIGO-India will further improve source localization and sky coverage \citep{unnikrishnan2013}.

Looking ahead to the 2030s, the field will be transformed by the advent of third-generation (3G) detectors such as the Einstein Telescope (ET) in Europe and Cosmic Explorer (CE) in the United States. These facilities will deliver an order-of-magnitude improvement in strain sensitivity, extending the detection horizon for compact binary coalescences to cosmological distances \citep[$z \sim 3$ for BNSs and $z \sim 20$ for BBH,][]{2023JCAP...07..068B} and increasing the detection rate of BNS mergers to more than $10^4$ per year \citep{2022A&A...665A..97R,2025A&A...697A..36L,2025arXiv250300116C,2026JCAP...03..081A}. 

The era of ET will benefit from major advances across the EM spectrum. In the optical and near-infrared, the Vera C.~Rubin Observatory \citep{ivezic2019,andreoni2022} will provide deep, wide-field surveys capable of promptly identifying KNe associated with GW events. Spectroscopic follow-up will be enabled by next-generation large-aperture telescopes such as the Extremely Large Telescope \citep[ELT;][]{marconi2022} and the Wide-field Spectroscopic Telescope \citep[WST;][]{mainieri2024}. 
At higher energies, new X-ray missions such as \textit{NewAthena} \citep{nandra2013} will enhance our ability to detect and characterize transient counterparts. In the MeV–GeV range, proposed missions such as \textit{THESEUS} \citep{2021ExA....52..183A} and modular satellite constellations such as \textit{HERMES} \citep{fiore2020,ghirlanda2024} promise to revolutionize high-energy transient monitoring through flexible, rapid-response architectures. Finally, in the very-high-energy (VHE, $>100$ GeV) regime, the Cherenkov Telescope Array Observatory \citep[CTAO;][]{CTA2019} will provide unprecedented sensitivity to non-thermal emission from the most extreme astrophysical events \citep[see][for an extensive overview of gamma-ray synergy opportunities for the SKAO]{Castignani01.2026.SKA}.
In the radio domain, the overlap between ET and SKAO (AA4) will deliver a powerful multi-messenger combination, providing the sensitivity and angular resolution for studying GRB afterglows and late-time emission from merger remnants. A complementary high-frequency view may be offered by the next-generation VLA, as outlined for instance by \citet{2017arXiv170908512L}.

\subsection{SKAO contribution to the detection and identification of GW counterparts}\label{sect2a}
Between the current observing runs of the LVK network and the deployment of 3G detectors, incremental upgrades to the existing 2G instruments are expected to extend the BNS detection horizon and moderately increase the number of observed mergers. These improvements will be valuable, but the truly transformative change will come with the arrival ET and CE, which will deliver more than an order-of-magnitude increase in strain sensitivity relative to the current network. This leap in performance will expand the accessible volume by several orders of magnitude and enable detection rates far beyond what can be achieved with upgraded 2G instruments.
In this section, we describe the expected radio afterglow emission from GRBs produced by BNS mergers detectable with 3G GW detectors, and we estimate the number of such events that SKAO may observe. We base our analysis on the BNS population model of \cite{2025arXiv250300116C}, calibrated on current GW and Galactic BNS observations and consistent with the latest constraints on the local merger rate. To explore the impact of future GW facilities, we consider three detector network configurations:
\begin{itemize}
    \item ET2L, two 15-km L-shaped interferometers in Sardinia and the Meuse–Rhine region;
    \item ET$\Delta$, the 10-km triangular ET design located in Sardinia;
    \item ET2L+2CE, an optimistic network combining ET2L with two Cosmic Explorer detectors (40 and 20 km).
\end{itemize}
 Radio afterglow light curves are computed at the central frequencies of SKAO bands 1 (0.80 GHz), 2 (1.31 GHz), 5a (6.55 GHz), and 5b (11.85 GHz) over timescales from 0.1 to 1000 days.

\begin{figure}[h]
    \centering
	\includegraphics[width=0.8\columnwidth]{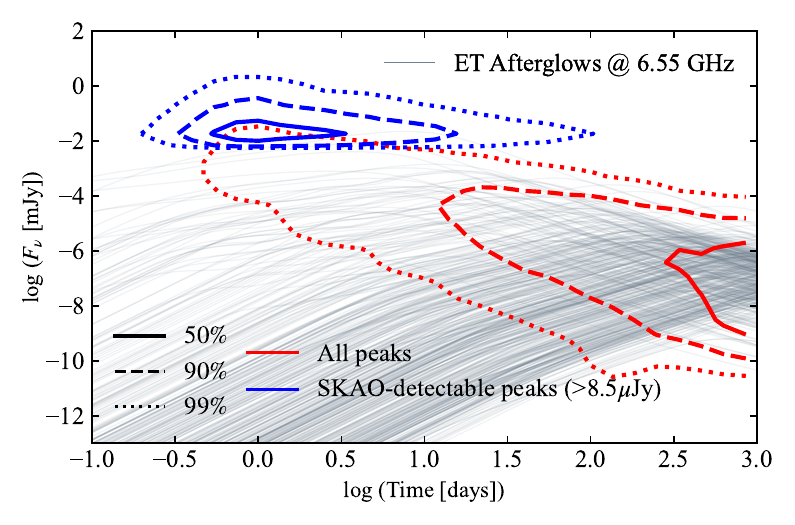}
    \caption{Random sample of GRB afterglow light curves (grey lines) at $6.55$ GHz for ET-detectable BNS mergers, assuming the ET2L network. The red contours enclose 50\%, 90\%, and 99\% of the peak-flux density distribution of the full population (solid, dashed, and dotted lines). The blue contours show the corresponding distribution for the subset of afterglows with peak flux densities above the SKAO–AA4 10-min $5\sigma$ sensitivity (8.5 $\mu$Jy in Band5A), highlighting the characteristic peak times of SKAO-detectable events.}
    \label{fig:after_radio_ska}
\end{figure}

Figure~\ref{fig:after_radio_ska} shows a sample of GRB afterglow light curves (grey lines) at 6.55 GHz for GW-detectable\footnote{With a GW signal-to-noise ratio larger than 12.} BNS mergers, assuming the ET2L detector network configuration. The red contours enclose 50\% (solid), 90\% (dashed), and 99\% (dotted) of the peak-flux density distribution of the full population. The blue contours show the subset of peak flux densities that exceed the 10-min SKAO-AA4 sensitivity $5\sigma$ (8.5 $\mu$Jy).

\begin{figure}[h]
    \centering
	\includegraphics[width=0.8\columnwidth]{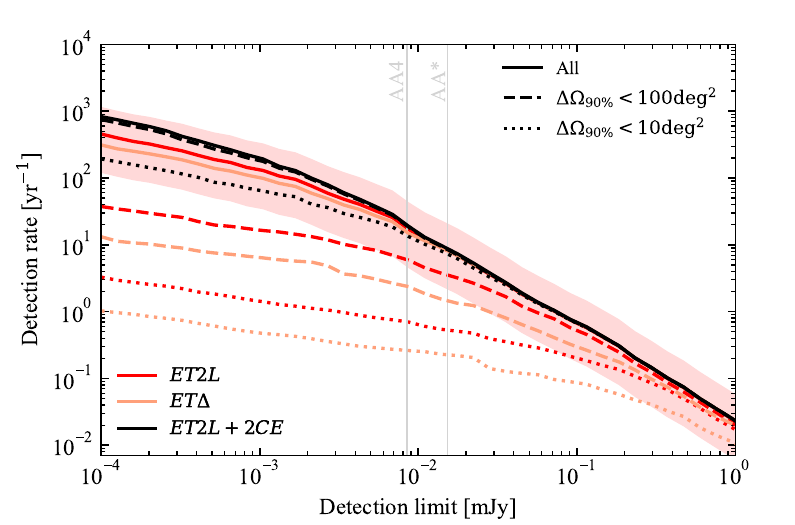}
    \caption{Joint GW and GRB afterglow at 6.55 GHz detection rate as a function of the EM detection limit threshold for BNSs. We indicate in red, light red and black the ET2L, ET$\Delta$ and ET2L+2CE configurations. The solid line indicates all the detectable binaries, the dashed and dotted lines the detectable binaries with a sky localization $\Delta\Omega_{{\rm 90}\%}<100\mathrm{deg}^2$ and the ones with $\Delta\Omega_{{\rm 90}\%}<10\mathrm{deg}^2$, respectively. The two grey vertical lines indicate the AA4 and AA* sensitivities at 5$\sigma$.}
    \label{fig:rate_after}
\end{figure}

Taking into account the full population, about 50\% of the afterglows peak at times larger than 100 days with flux densities below $10^{-4}$ mJy. This reflects the dominance of off-axis events within the ET detection horizon, which produce radio emission that is too faint to be observed even with future radio facilities. However, a bright tail peaks at earlier times and higher flux densities. Although SKAO will detect only a minority of all BNS afterglows, the high ET detection rate ($>10^4\,\mathrm{yr^{-1}}$) ensures a significant number of SKAO detections (see Figure \ref{fig:rate_after}).

When focusing on the subset of afterglows whose peak flux density is above the SKAO–AA4 sensitivity, we recover a very different temporal behaviour. Roughly 50\% of SKAO-detectable peaks occur between $\sim12$ hours and $\sim3$ days after merger, and 90\% of them occur within $\sim30$ days. Only a small fraction peak earlier than 12 hours or later than 1 month. This indicates that extremely rapid ToO response (minutes) or very long-term rapid ToO readiness (months) are less critical for this science case. Instead, SKAO follow-up capabilities on timescales of hours to days are the most relevant for capturing the majority of detectable radio peaks.

To provide quantitative estimates of SKAO detection rates for BNS afterglows, Figure~\ref{fig:rate_after} presents the number of detectable events as a function of the flux density threshold (in mJy) for the ET2L, ET$\Delta$, and ET2L+2CE detector configurations, shown respectively in red, light red, and black. The curves correspond to GW+GRB afterglows at 6.55 GHz for all binaries (solid line), binaries with localization $\Delta\Omega_{90\%} < 100\,\mathrm{deg}^2$ (dashed line), and  $\Delta\Omega_{90\%} < 10\,\mathrm{deg}^2$ (dotted line). The two vertical grey lines mark the 5$\sigma$ sensitivities of SKAO configurations AA4 (8.5 $\mu$Jy) and AA* (15.25 $\mu$Jy) with an exposure of 10 minutes. The detection rate for any given flux density limit can thus be inferred directly from the plot. 
Assuming the AA4 (AA*) sensitivity, we obtain a detection rate of $18^{+28}_{-13}$ ($8.4^{+13.0}_{-6.2}$) events per year\footnote{At 0.80 GHz, 1.31 GHz and 11.85 GHz the rates are $3.9^{+6.0}_{-2.9}$ ($1.3^{+2.1}_{-1.0}$), $10.5^{+16.2}_{-7.8}$ ($3.6^{+5.6}_{-2.7}$), $12.3^{+19.0}_{-9.1}$ ($6.2^{+9.6}_{-4.6}$), respectively.} for the ET2L configuration. Restricting to sources localized within $100\,\mathrm{deg}^2$ and $10\,\mathrm{deg}^2$, the corresponding rates are $6.1^{+9.4}_{-4.5}$ ($3.6^{+5.6}_{-2.7}$) and $0.71^{+1.10}_{-0.52}$ ($0.53^{+0.81}_{-0.39}$) events per year, respectively. For smaller localization areas, longer exposure times would allow deeper sensitivities, thus increasing the detection rate. It is important to note that these values represent potentially detectable events, as we assume all GW-detected binaries with afterglows exceeding the flux density threshold at any epoch. This is a simplified estimate, since instrumental and operational factors—such as field of view, visibility constraints, and duty cycle—are not included here and will be addressed in future work.

A comparison with existing radio facilities highlights the unique contribution of SKAO–AA4 in the ET era. When sensitivities are normalised to a 10-minute integration, SKAO–AA4 is approximately 4 times more sensitive than MeerKAT \citep{2020ApJ...888...61M} and has a wider frequency coverage, 20 times more sensitive than the VLA \citep{2024ApJ...972...89J} and has a larger field of view, and nearly 200 times more sensitive than ASKAP \cite[][although ASKAP Phased Array Feeds make up for the loss in sensitivity when it comes to surveying large sky areas]{2022MNRAS.510.3794D}. This improvement translates directly into detection rates: replacing SKAO with MeerKAT reduces the number of detectable BNS afterglows by more than an order of magnitude, the VLA yields rates lower by almost two orders of magnitude, and ASKAP drops by over three orders of magnitude.

\subsection{Characterisation of identified counterparts}\label{sect2b}

As argued by \citet{2021MNRAS.505.2647D}, sensitive multi-frequency radio observations not only aid in the localisation of BNS mergers, as discussed in the previous section, but also play a key role in characterising the counterparts of GW events.
This was the case for GW\,170817, whose first detection at radio frequencies came 16 days after the burst \citep{2017Sci...358.1579H}, remarkably late compared to the typical rise time for radio afterglows of short and long GRBs \citep{2012ApJ...746..156C,2025ApJ...982...42S}. The observed radio emission could initially be  explained by either a slightly off-axis, collimated, relativistic jet or a mildly relativistic cocoon, and later observations were required to discriminate between the two \citep{2018ApJ...863L..18A,2018Natur.554..207M,2018Natur.561..355M,2019Sci...363..968G,2019MNRAS.489.1919T}.  Multi-frequency, multi-epoch, multi-scale (including VLBI) observations over the following months favoured a scenario with a successful structured jet and provided constraints also on the density profile of the circumbinary medium.  

While varying with instrument, observing frequency, and integration time, the typical image r.m.s.\ noise of the observations of GW\,170817 was of order $\sim10\mu$Jy\,beam$^{-1}$. Current estimates for SKA-MID $1\sigma$ sensitivity, assuming the AA* configuration, 10-minute integration time, and Briggs image weighting, are 6.1, 3.1, 3.7 $\mu$Jy\,beam$^{-1}$ in Bands 2, 5a, 5b, respectively. In the AA4 configuration, the values further improve to 2.8, 1.7, 2.1 $\mu$Jy\,beam$^{-1}$. Overall, studies similar to those carried out for GW\,170817 would become feasible to distances that are several times as large, implying a volume and an event rate at least one magnitude higher. The added sensitivity of SKA-MID as an element of a VLBI array will increase the number of sources for which the signal-to-noise ratio will be large enough to measure the effects of apparently superluminal motion or expansion, as detailed in the Chapter by \citet[][see also \citealt{2020MNRAS.494.2449D}]{Giarratana01.2026.SKA}. 

At the same time, the increased sensitivity of SKAO will be transformational for advanced studies of events at a distance comparable to that of GW\,170817: this includes the possibility of revealing polarised emission \citep{2018ApJ...861L..10C}, thus probing the magnetic field configuration, and of revealing the expected late-time radio emission from the deceleration of the KN ejecta, which has so far remained undetected (see next section).

\subsection{Kilonova radio remnants} \label{sect2c}

In addition to the radio afterglow emission from relativistic jets, BNS mergers are also expected to produce late-time radio emission powered by the interaction between the sub-relativistic KN ejecta and the surrounding medium. Following the thermal KN phase, the merger ejecta continue to expand into the surrounding interstellar medium (ISM). The fastest ejecta move at highly supersonic velocities and create a forward shock in the ISM. Meanwhile, the slower material eventually encounters a reverse shock, transferring its kinetic energy into the shocked region and helping to power the non-thermal emission that develops.

Electrons accelerated at the shock front produce synchrotron radiation as they interact with the local magnetic fields. This emission emerges predominantly in the radio band and evolves on extremely long timescales—typically years to decades after the merger. In the literature, this component is sometimes referred to as a “radio flare” \citep{nakar2011} or, given its slow evolution and analogy to supernova remnants, as a “kilonova radio remnant” \citep{barbieri2019}. 

The expected strength of this radio component has been modeled for both neutron-star mergers and black-hole–neutron-star mergers \citep{hotokezaka2015}. Depending on physical conditions such as the ejecta mass, velocity distribution, and especially the ambient ISM density, the peak flux density could reach the mJy level. However, these predictions carry significant uncertainties, and the detectability of such emission remains an open observational challenge.

The best opportunity to detect this component so far has been provided by GW170817. At $\sim900$ days, observations with the Chandra X-ray Observatory revealed X-ray emission in excess of the declining jet afterglow, possibly hinting at the emergence of a kilonova afterglow \citep{hajela_2022}. Deep radio follow-up with the Karl G. Jansky Very Large Array, extending to 3.5–4.5 years post-merger yielded no significant rebrightening and placed $\mu$Jy-level upper limits on any kilonova radio remnant component \citep{2021ApJ...914L..20B,balasubramanian_2022}. These constraints already disfavour models with substantial energy in a fast ejecta tail, although continued monitoring on decadal timescales remains essential.

The search for KN radio remnants has been performed also for other KN detected independently of GW events, but without success (e.g.: \citealt{2026ApJ...998...93S} for the KN associated with GRB\,211211A).

Given the long timescales and the expected faintness of most KN remnant signals, their identification and characterization will require deep, high-sensitivity radio observations. Long-lived supermassive neutron star remnants of neutron star mergers are expected to produce extremely bright KN remnants. However, they have not been detected so far, limiting their fraction to less than 30\% of the KN remnant population \citep{2025A&A...693A.108K}.  The $\mu$Jy-level continuum capabilities of SKAO, together with its long-term monitoring stability, make it uniquely suited to also probe the faint emission regime and to test the predicted diversity of KN radio remnants in the ET era.

\section{The contribution of the SKAO in lack of a GW trigger: afterglow studies of long and short GRBs}\label{s.3} 

The SKAO will be able to provide crucial contributions to address a lot of unanswered questions in the study of GRB afterglows, regardless of their physical origin (merger or collapsar) and detection method (GW, high energy, or other).

The study of cosmological GRBs is a rapidly evolving field: in the last 5 years it has been demonstrated, using e.g.\ the VLA and SKA precursor MeerKAT, that the radio counterparts of both short and long GRBs show complex light curve morphology \citep{2024ApJ...970..139S,2025ApJ...994....5A}. Despite these huge strides, observing campaigns are still sensitivity limited and rarely can provide more than a few temporal or spectral points per event \citep[e.g.][]{2025GCN.39520....1G}. On a population level, SKA will let us probe the full range of the radio luminosity function of long and short GRBs at multiple frequencies, tracing the evolution of the peak of the emission, sampling the thick-to-thin transition of the spectrum, and converting this information to quantities such as the electron energy distribution and magnetic field.  Studying systematically all the afterglows of GRBs down to $\mu$Jy-level sensitivity on different time scales will also probe whether there really is a dichotomy between radio-bright and radio-dark GRBs \citep[][and references therein]{2023MNRAS.520.5764C}.

Furthermore, a sensitive multi-frequency instrument like the SKA will also let us study each individual GRB afterglow in great detail, to understand the diversity of these events and the underlying physics as they deviate from the most simplistic models. This is a recently opened avenue, presently in the spotlight, following the detection of outstanding events such as GRB~221009A \citep[see e.g.][]{2023ApJ...946L..23L,2024MNRAS.533.4435R}. Full characterisation of GRB afterglows with radio telescopes requires multi-frequency observations across a wide time range. At present, rapid response mode observations are carried out with the Murchison Widefield Array \citep{2021PASA...38...26A} and the Australia Telescope Compact Array \citep{2021MNRAS.503.4372A,2026arXiv260319047C} shedding light on early time emission as the jet begins to decelerate and the reverse shock onset occurs; the SKAO will be vital in the search for coherent, prompt radio signals associated with the initial event and central
engine and the early synchrotron emission generated by the outflows \citep{AlexAndersson01.2026.SKA}.  
At very late times, the increased sensitivity will enable the tracking of the afterglow deep into the non-relativistic regime when the outflow is full spherical, allowing for calorimetry calculations and constraining of the circumstellar density profile. Currently, it is only possible to study the brightest events out to very late times introducing bias into the resulting sample. A more sensitive telescope will let us measure the total energy budget for a more complete sample of events. 

SKA's incredible sensitivity will be vital also in the study of objects that cross the traditional long-short GRB classification boundary, i.e.\ short GRBs from collapsars and long GRBs from BNS mergers. To date there have only been a handful of these events discovered but in each case the radio counterparts have been at/below the sensitivity limit of the present generation of radio telescopes \citep{2021MNRAS.503.2966R, 2024Natur.626..737L}. In luminosity space, long GRB from BNS systems sit over an order of magnitude below the rest of the radio-detected short GRB population. Targeting long GRB BNS mergers with the SKA will allow us to (1) confidently detect the afterglows and (2) investigate any difference between the afterglows of long GRB BNS mergers and `regular' short GRB events. 

Polarisation observations will be useful to provide orthogonal information on the magnetisation and geometry of the jet. Despite the wealth of understanding that can be gained from polarisation calibrated observations, only a single detection at SKA frequency has been reported so far \citep{2026arXiv260427480C,2018ApJ...861L..10C}; mm-wavelength polarised emission just below 1\% was detected in the afterglow of GRB 190114C few hours after the burst \citep{2019ApJ...878L..26L}. Taking into account the temporal decrease of the flux density by the time it becomes detectable at the SKA frequencies, detecting a comparable polarisation fraction will require $\mu$Jy sensitivity.

Furthermore, by operating as an element of a VLBI array, SKAO will enable the characterisation of the structure of the jet and the determination of independent constraints on the viewing
angle and jet opening angle. This is the subject of a dedicated Chapter in this book \citep{Giarratana01.2026.SKA} to which we refer the reader, also pointing out the importance of baselines within the African continent \citep{Bempong-Manful01.2026.SKA} for improved performance of such an array.

Last but not least, independently of GW triggers, SKAO will enable the detection of GRB off-axis afterglows, especially those of long GRBs. This is a line of research that is already delivering promising results through the search for relativistic jets associated with Ic-BL SNe \citep[][]{2023ApJ...953..179C,2025ApJ...995...61S}; it will require the SKAO sensitivity to deliver systematic conclusions.  Off-axis events can be discovered through transient commensal searches in SKAO surveys (see also \citealt{AlexAndersson01.2026.SKA}) following the methodology recently developed for surveys with the Australian SKA Pathfinder \citep{2023MNRAS.523.4029L,2026ApJ..1000..118G}; in addition, SKAO can carry out follow-up observations of candidate off-axis afterglows detected in optical surveys, such as LSST, or by current (e.g.:\ Einstein Probe, SVOM) and future space missions (e.g.: THESEUS; \citealt{2021ExA....52..183A}) with large FoV high-energy monitors sensitive to the X-rays.
SKAO surveys reaching $\mu$Jy limits are expected to detect of the order of $\sim 0.5$~deg$^{-2}$~yr$^{-1}$ orphan afterglows \citep{2014PASA...31...22G}.
Rubin LSST single visit depth ($r\sim24.7$) should be able to detect roughly 50 off-axis GRB afterglow per year \citep{2023PASP..135j5002H}. 
The correct identification of such events and/or their characterization require SKAO long-term campaign of observations over months/years, but, considering the expected faintness of off-axis afterglows, SKAO is one of the few facilities allowing their identification and study.

\vspace{0.7 cm}

\begin{tcolorbox}[colback=blue!5!white,colframe=blue!75!black,title=\bf Executive Summary]
   SKAO will be transformational for the study of the electromagnetic counterparts of GW detections of binary neutron star merger events, but also for the study of GRBs and their jets, independently of a GW detection. Thanks to SKAO it will be possible to:

\begin{itemize}
\vspace{0.2 cm} 
\item detect and study a population of on- and off-axis afterglows of GW BNS events;
\item detect and study for the first time KN remnants of BNS mergers;
\item obtain multi-frequency light curves (multi-epoch spectra) for the bulk of the population of long- and short-GRBs (and not some bright sparse event);
\item study the structure of GRB jets thanks to the gathering of a sample of off-axis GRB afterglows; 
\item scale from a single case study to a larger sample the set of GRB radio polarisation measurements, capital to study the magnetisation and geometry of the jet.
\end{itemize}

\vspace{0.2 cm}
These observations will be obtained in synergies with the major current and future facilities detecting GW, GRBs, and their (off-axis) afterglows, such has LVK, ET, CE, ELT, NewATHENA, VRO, Swift, Fermi, SVOM, EP, CTAO, ELT, and  through commensal searches in SKAO surveys. 
\end{tcolorbox} 

\bibliographystyle{abbrvnat-maxbibnames4}
\bibliography{chapter} 

\end{document}